\documentclass{jaa}
\usepackage{fix-cm}
\usepackage[T1]{fontenc}
\usepackage{ae,aecompl}
\usepackage[normalem]{ulem}
\usepackage{graphicx}	
\usepackage{amsmath, mathtools, xfrac}	
\usepackage{amssymb, enumitem}	
\usepackage{longtable, multicol}
\usepackage{rotate, setspace}
\usepackage{lscape}
\usepackage{natbib, hyperref}
\usepackage{latexsym,verbatim,xcolor}
\newcommand{\apj}[1]{ApJ, }
\newcommand{\jcp}[1]{J. Chem. Phys., }
\newcommand{\mnras}[1]{MNRAS, }
\newcommand{\aj}[1]{AJ, }
\newcommand{\apjs}[1]{ApJS, }
\newcommand{\apjl}[1]{ApJ Letter, }
\newcommand{\aap}[1]{A\&A, }
\newcommand{\aaps}[1]{A\&A Suppl. Series, }
\newcommand{\araa}[1]{Annu. Rev. A\&A, }
\newcommand{\aaas}[1]{A\&AS, }
\newcommand{\apss}[1]{Ap\&SS }
\newcommand{\bain}[1]{Bul. of the Astron. Inst. of the Netherlands,}
\newcommand{\planss}[1]{Planetary and Space Science,}
\newcommand{\nat}[1]{Nature,}
\newcommand{\actaa}[1]{Acta Astronomica,}
\newcommand{\aapr}[1]{The Astronomy and Astrophysics Review,}
\newcommand{\memsai}[1]{Memorie della Societa Astronomica Italiana,}

\begin{document}\sloppy
\title{Deeply Comprehensive Astrometric, Photometric, and Kinematic Studies of the Three OCSN Open Clusters with $Gaia$ DR3} 

\author{W. H. Elsanhoury\textsuperscript{1*}, Haroon A. A\textsuperscript{2}, E. A. Elkholy\textsuperscript{1,3}, D. C. \c{C}ınar\textsuperscript{4}}
\affilOne{\textsuperscript{1}Department of Physics, College of Science, Northern Border University, Arar, Saudi Arabia.\\}
\affilTwo{\textsuperscript{2}Astronomy and Space Science Department, Faculty of Science, King Abdulaziz University, Jeddah, Kingdom of Saudi Arabia.\\}
\affilThree{\textsuperscript{3}Astronomy Department, National Research Institute of Astronomy and Geophysics (NRIAG), 11421, Helwan, Cairo, Egypt.\\} 

\affilFour{\textsuperscript{4}Istanbul University, Institute of Graduate Studies in Science, Programme of Astronomy and Space Sciences, 34116, Beyazıt, Istanbul, Turkey.}

\twocolumn[{

\maketitle

\corres{elsanhoury@nbu.edu.sa, elsanhoury@nriag.sci.eg}

\msinfo{XX XX XXXX}{XX XX XXXX}

\begin{abstract}

In this study, we considered the optical wavelength of $Gaia$ Data Release 3 (DR3) to analyze poorly studied three newly open star clusters namely OCSN 203, OCSN 213, and OCSN 244 clusters with {\sc ASteCA} code. Here, we identified candidates of 227, 200, and 551 with highly probable ($P \geq 50\%$) members. Fitting King’s profile within radial density profiles allows us to estimate inner stellar structures like core (0.190 $\le r_{\rm c}$ (pc) $\le$ 1.284) and the limiting (0.327 $\le r_{\rm cl}$ (pc) $\le 1.302$) radii. Constructing color-magnitude diagrams (CMD) fitted with suitable log age (yr) between ($\log t$; 6.52 - 7.05) and metallicities ($Z$; 0.01308 – 0.01413) isochrones. Therefore, the estimated photometric parameters with CMDs, reflect the heliocentric distances are 332 $\pm$ 18, 529 $\pm$ 23, and 506 $\pm$ 23 (pc) for OCSN 203, OCSN 213, and OCSN 244, respectively. Furthermore, the collective mass ($M_{\rm C}$) in solar mass units calculated with MLR as 67 $\pm$ 8.19, 91 $\pm$ 9.54, and 353 $\pm$ 18.79. Additionally, LF determined that the mean absolute magnitudes are 9.54 $\pm$ 3.09,	8.52 $\pm$ 2.92, and 7.60 $\pm$ 2.76 for these clusters, respectively. The overall mass function reflects the slopes ($\alpha$) for Salpeter within the uncertainty are ($\alpha_{\rm OCSN~203}$ = 2.41 $\pm$ 0.06), ($\alpha_{\rm OCSN~213}$ = 2.13 $\pm$ 0.07), and ($\alpha_{\rm OCSN~244}$ = 2.28 $\pm$ 0.07). The results of this study, which employed a dynamical analysis over varying timescales, indicate that OCSN 203 and OCSN 244 are clusters that have undergone significant relaxation, with a dynamical evolution parameter ($\tau$) that is much greater than one. In contrast, OCSN 213 exhibits characteristics of a non-relaxed cluster. A kinematic analysis of these open clusters was carried out, encompassing aspects of their apex position ($A_{\rm o},~D_{\rm o}$) using the AD diagrams. Therefore, the numerical convergent point coordinates are 76$^o$.77 $\pm$ 0$^o$.01, -0$^o$.23 $\pm$ 0$^o$.00 (OCSN 203), 85$^o$.71 $\pm$ 0$^o$.11, -9$^o$.63 $\pm$ 0$^o$.03 (OCSN 213), and 88$^o$.19 $\pm$ 0$^o$.11, -4$^o$.04 $\pm$ 0$^o$.01 (OCSN 244). At the end, we found that the three OCSN clusters are young stellar disc members using dynamic orbit parameters.

\end{abstract}


}]

\doinum{12.3456/s78910-011-012-3}
\artcitid{\#\#\#\#}
\volnum{000}
\year{2021}
\pgrange{1--}
\setcounter{page}{1}
\lp{\#}

\section{Introduction}
An open cluster (OC) is defined as a collection of stars that have formed from the same molecular cloud and are gravitationally bound together. They represent essential elements of galaxies, as they contain the fundamental building blocks of stellar systems \citep{Lada03,de_Wit2005,Portegies10}. OCs are fascinating possibilities because they provide essential information on star kinematics, evolution, and structure of the Galactic disk. Most star births are thought to occur in OCs, which are generated within giant molecular clouds (GMCs) \citep{Yadav11}. Thus, their study will improve our knowledge of star evolution and increase our understanding of the Milky Way’s (MW) evolution and structural makeup. To identify clusters, the researchers employed a variety of techniques, several of which were predicated on an unsupervised machine-learning clustering algorithm ({\sc UPMASK}; \cite{Krone-Martins14}). In a prior paper, \cite{He21} employed this approach to search $Gaia$ DR2 for OCs within 20$^o$ of the Galactic planes; they found 986 new OC candidates, 74 of which have been reported previously. Based on this information, He and colleagues \citep{He22-a,He22-b} went over the remaining candidates again and used $Gaia$ EDR3 to verify which of them were true new OCs. Utilising data from the $Gaia$ DR3, \citet{Qin23}\footnote{\url{https://vizier.cds.unistra.fr/viz-bin/VizieR?-source=J/ApJS/265/12}} conducted a comprehensive and systematic search for OCs at Galactic latitudes of $|b| \le 30^{o}$ within 500 pc of the solar neighbourhood. This involved varying the size of the slice boxes in different distance grids. A total of 324 OCs were identified by employing the clustering algorithms {\sc pyUPMASK} and {\sc HDSBSCAN} (Hierarchical Density-Based Spatial Clustering of Applications with Noise) \citep{2017JOSS....2..205M}. The findings include 101 previously unreported additional clusters and provided their Open Cluster of Solar Neighborhood (OCSN), boosting the OC census within 500 pc.

Utilising the OCSN catalogue, \citet{Qin23} estimated the membership probabilities of stars within each cluster and subsequently calculated key parameters, including the mean positions, proper motions, parallaxes and structural properties. Furthermore, through Gaussian fitting applied to the $Gaia$ DR3 data, they determined the mean radial velocities of the OCs. Furthermore, a visual isochrone fitting procedure was conducted to estimate the reddening values, distance moduli, and cluster ages based on the distribution of member stars observed on the colour-magnitude diagrams (CMDs).

This study is considered the second continuous article intended to provide a comprehensive characteristic of new hunting OCs by \cite{Qin21,Qin23} whereas our first one is achieved before for OCs \citep{Elsanhoury22}. Our new project considered new three OCs; OCSN 203, OCSN 213, and OCSN 244. The first two clusters appear in the recent catalogue by \citep{Hunt24} known as IC 348 and L 1641S. In this analysis, we utilized the Automated Stellar Cluster Analysis ({\sc ASteCA}) code \citep{Perren15}, which was chosen due to its robust functionality in deriving key cluster parameters with minimal user input. This tool is specifically designed to handle both positional and photometric data, making it particularly suitable for our objectives. One of the main advantages of {\sc ASteCA}, compared to other available codes, is its open-source accessibility, allowing it to be publicly used and tested across various OCs. This wide applicability and thorough validation make it a reliable choice for conducting comprehensive cluster analyses (e.g., \cite{Perren15,Perren20,Elsanhoury22}). Future observations and studies are needed to characterise the properties of the three clusters, the authors conclude.

The following is a description of the structure of this paper: In Section \S~\ref{sec2}, we outline the data preparation process is outlined detailed with the specific criteria used for data selection and the creation of relevant subsets. The internal structural analysis of the clusters is explored in Section \S~\ref{sec3}. Section \S~\ref{sec4} focuses on the identification and selection of probable cluster members, which are then employed in the construction of CMDs. In Section \S~\ref{sec5}, we present the luminosity and mass functions, including the mass-luminosity relation. Section \S~\ref{sec6} discusses the clusters' dynamical evolution times, kinematic properties, and orbital dynamics. Finally, Section \S~\ref{sec7} provides the concluding remarks.

\section{Gaia Data DR3}
\label{sec2}
The most up-to-date astrometric data are provided by the $Gaia$ Mission Collaboration's Data Release 3 ($Gaia$ DR3; \cite{2021A&A...649A...1G})\footnote{\url{https://cdsarc.cds.unistra.fr/viz-bin/cat/I/355}}. Since the European Space Agency (ESA) launched the $Gaia$ mission, it has revolutionized astronomy, offering a new era of precise astrometric measurements. $Gaia$ DR3 comprises five-parameter astrometry for approximately 1.8 billion sources, with a limiting magnitude of $G = 21$ mag. These parameters include the celestial coordinates ($\alpha$, $\delta$), parallaxes ($\varpi$), and proper motions in both directions ($\mu_\alpha\cos\delta,~\mu_\delta$). Additionally, $Gaia$ DR3 provides photometric data across three broadband filters: the $G$ band (330–1050 nm), the Blue Prism ($G_{\rm BP}$: 330–680 nm), and the Red Prism ($G_{\rm RP}$: 630–1050 nm) for sources brighter than 21 mag \citep{2018A&A...617A.138W}. The uncertainties in photometry magnitudes $\sim$ 0.30 for $G \le 20$ mag with three photometric bands $G,~G_{\rm BP},$ and $G_{\rm RP}$.

The proper motions uncertainties range from 0.02 to 0.03 mas yr$^{-1}$ (at $G \le 15$ mag), 0.07 mas yr$^{-1}$ (at $G$ $\sim$ 17 mag), 0.50 mas yr$^{-1}$ (at $G$ $\sim$ 20 mag) and 1.40 mas yr$^{-1}$ (at $G$ = 21). The parallax values have errors of 0.02 to 0.03 mas for sources with $G \le 15$ mag, $\sim$ 0.07 mas for sources with $G = 17$, $\sim$ 0.50 mas at $G=20$ mag, and $\sim$ 1.30 mas for sources with $G = 21$ mag. Data on the radial velocity ($V_{\rm r}$; km s$^{-1}$) for approximately 7 million stars from $Gaia$ DR2 are added to the DR3 \citep{2021A&A...649A...1G}\footnote{\url{https://cdsarc.cds.unistra.fr/viz-bin/cat/I/350}}. A notable advancement in the precision of astrometric parameters can be observed, with the factor of 2.00 in proper motion accuracy and a factor of roughly 1.50 in parallax accuracy representing a significant distinction between the $Gaia$ DR3 and $Gaia$ DR2 datasets. A reduction in the astrometric errors by a factor of 2.50 for the proper motion and by a range of 30–40$\%$ for the parallax has also been observed.

\begin{figure}
\begin{center}
\includegraphics[width=0.85\linewidth]{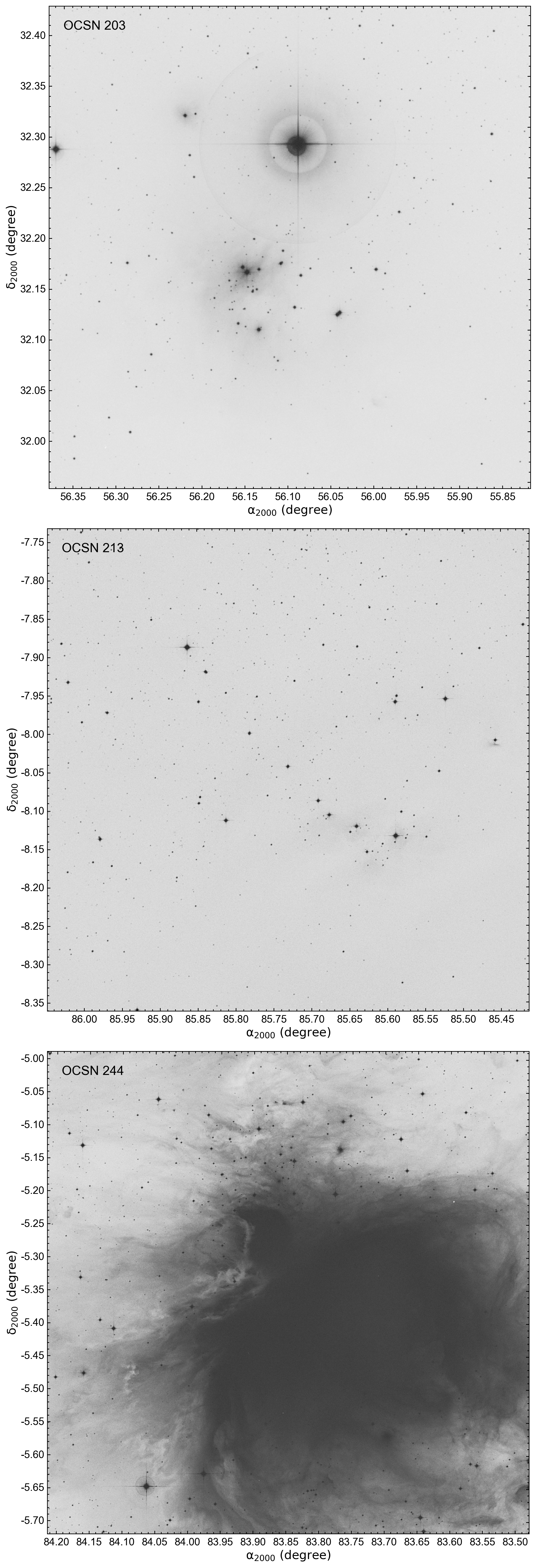}
\caption {Identification maps of OCSN 203, OCSN 213, and OCSN 244.}
\label{fig1}
\end{center}
\end{figure}

In this analysis, we have employed the fundamental parameters of the three selected OCSNs, as inferred by \cite{Qin23}, as listed in Table \ref{tab1}. Identification maps of OCSN 203, OCSN 213, and OCSN 244 shown in Figure \ref{fig1}, obtained from SIMBAD Astronomical Database - CDS (Strasbourg)\footnote{https://simbad.u-strasbg.fr/simbad/}. The M43 Orion nebula in front of the OSCN 244 open cluster, projected in Figure \ref{fig1}, did not affect the analyses. This study uses $Gaia$ DR3 data, which is capable of separating objects at this distance. In order to commence the calculations, it is first necessary to download the data that has been employed, including the angular distances (in arcmin) from the center via $Gaia$ DR3 and both right ascension ($\alpha_{2016}$) and declination ($\delta_{2016}$). A star count was conducted on extracted data within a 40-arcminutes radius for OCSN 203, OCSN 213, and OCSN 244.

\begin{table}
\caption{Initial parameters of three OCSNs under investigation by \citet{Qin23}.}
\label{tab1}
\scriptsize
\setlength{\tabcolsep}{4pt}
\centering
\begin{tabular}{lccc} \hline 
{\bf Parameters}& {\bf OCSN 203} & {\bf OCSN 213}	& {\bf OCSN 244}  \\
\hline
\hline
$\alpha$	& 03$^h$44$^m$ 33$^s$.60 & 05$^h$42$^m$ 23$^s$.60 & 05$^h$35$^m$ 16$^s$.80 \\[0.8 ex]
$\delta$	& 32$^o$09'00''.00 & -08$^o$07'48''.00 & -05$^o$ 23'24''.00 \\[0.8 ex]
$l$	        & 160$^o$.7978 & 212$^o$.4613 & 209$^o$.0117 \\[0.8 ex]
$b$	        & -17$^o$.8134  & -18$^o$.9839 & -19$^o$.3830 \\[0.8 ex]
r$ _{\rm c}$ (arcmin)& 8.40 $\pm$ 0.00 &	17.40 $\pm$ 0.24	&	4.20 $\pm$ 0.00	\\[0.8 ex]
r$ _{\rm cl}$ (arcmin)& 37.80 $\pm$ 0.12 &	75.00 $\pm$ 0.54	&	20.40 $\pm$ 0.12	\\[0.8 ex]
log age (yr) & 6.25 &	6.70	&	6.65	\\[0.8 ex]
$E$($B$-$V$) & 0.11 &	0.48	&	0.53	\\[0.8 ex]
$\varpi$ (mas)& 3.17 $\pm$ 0.13 &	2.35 $\pm$ 0.07	&	2.49 $\pm$ 0.13	\\[0.8 ex]
$d_{\rm \varpi}$ (pc)& 316 $\pm$ 0.03 &	426 $\pm$ 0.03	&	402 $\pm$ 0.05	\\[0.8 ex]
$\mu_{\alpha} \cos{\delta}$ (mas yr$^{-1})$& 4.47 $\pm$ 0.55 &	0.10 $\pm$ 0.27	&	1.41 $\pm$ 0.78	\\[0.8 ex]
$\mu_{\delta}$ (mas yr$^{-1})$& -6.30 $\pm$ 0.57	&	-0.33 $\pm$ 0.32 & 0.35 $\pm$ 0.89 	\\[0.8 ex]
V$_{\rm r}$ (km s$^{-1}$)& 11.18 $\pm$ 15.97 &	15.67 $\pm$ 12.32	&	23.68 $\pm$	8.20\\[0.8 ex]
\hline
\end{tabular}
\end{table}

\section{Structural properties}
\label{sec3}
\subsection{\bf Cluster center determination}

Identifying the center of an OC is not as straightforward as it often is for globular clusters, where the core is more visually distinct. To accurately determine the center of the OCs in our study, we used the {\sc ASteCA} package, which employs a standard technique of assigning the cluster's central coordinates to the point of maximum spatial density. This point is identified by searching for the highest value of a two-dimensional Gaussian kernel density estimator (KDE) applied to the positional diagram of the cluster. Unlike visual estimations, this KDE-based approach provides a more objective and reliable method for center determination.

While there are only a few algorithms available for automatic OC center identification in the literature, various methods have been applied. For example, \cite{Bonatto07} used visual estimations as a starting point, refining them by searching for the point of maximum star density in XDSS7 images using a two-dimensional histogram. Similarly, \cite{Maciejewski07} utilized initial estimates from the DAML02 catalogue, refining them using a comparable histogram-based approach. \cite{Maia14} employed an iterative algorithm that starts with estimates of the cluster's center and radius (based on \cite{Bica08}) and refines these by averaging the positions of stars, weighted by their local stellar density.

These methods, however, depend on accurate initial estimates of the center and radius to function properly. In contrast, the {\sc ASteCA} package does not require initial values for the cluster center, though such values can be provided in a semi-automatic mode. Convergence is always guaranteed, which makes the process robust and reliable. Furthermore, {\sc ASteCA} overcomes the limitations of binning, as it calculates the KDE bandwidth using Scott's rule \citep{Scott92}. By simultaneously estimating the maximum density in both spatial dimensions, it minimizes deviations caused by densely populated fields. This approach is coordinate-system-independent, allowing it to be equally effective with positional data recorded in either pixels or degrees.

\begin{figure*}
\centering
\includegraphics[width=0.8\linewidth]{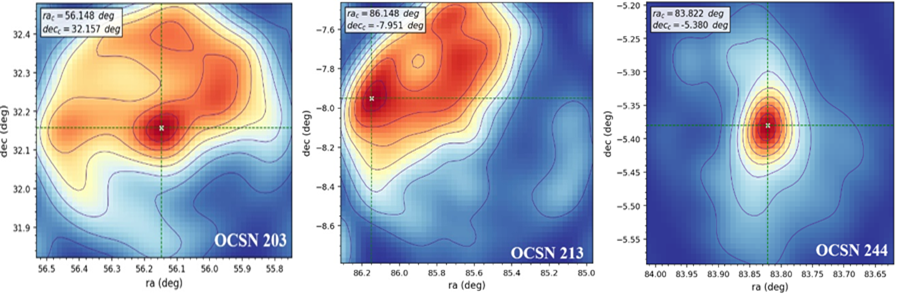}
\caption {Contour maps of OCSN clusters.}
\label{fig2}
\end{figure*}

Figure \ref{fig2} shows OCSN's contour maps for re-estimation of the center with Automated Stellar Cluster Analysis ({\sc ASteCA}) code \citep{Perren15}, where star counts tend to concentrate like the deepest point in the figure. The findings of new centers of the three clusters exhibit a modification in the right ascension direction like ($\Delta\alpha\le 2^m~04^{\rm s}.00$) and declination. ($\Delta\delta\le 10^m~52^{\rm s}.20$) with those adopted ones devoted by \cite{Qin23}, under or overestimated numerical values depending on the number of possible data used and the method of estimation. The revised estimated centers of the clusters in the equatorial ($\alpha,~\delta$) and Galactic ($l,~b$) coordinate systems are presented in Table \ref{tab2}.

\begin{figure}
\centering
\includegraphics[width=0.99\linewidth]{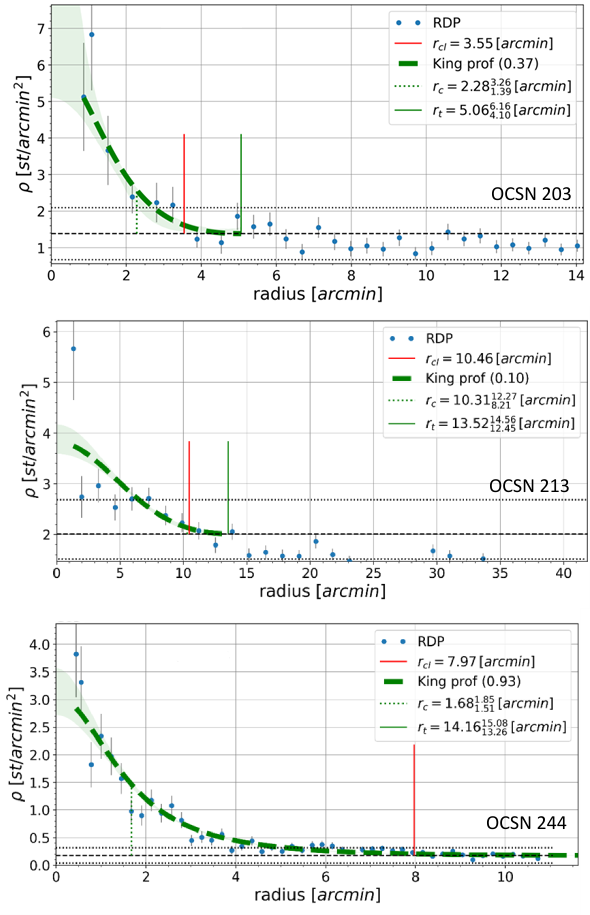}
\caption {RDPs for OCSN 203, 213, and 244 are shown in the upper, middle, and lower panels, respectively. The dashed blue lines represent the fit using King's density distribution model, while the horizontal dashed lines indicate the background field density ($\rho_{\rm bg}$; stars per arcmin$^{-2}$). The vertical solid red, green, and dotted green lines mark the cluster’s limiting radius ($r_{\rm cl}$), tidal radius ($r_{\rm t}$), and core radius ($r_{\rm c}$), respectively, all measured in arcminutes. These visualizations provide a clear comparison of the spatial structure and density distribution of each cluster.}
\label{fig3}
\end{figure}

We used the {\sc ASteCA} code to create three radial density profiles (RDPs), for each cluster and fitted each RDP with King’s profile \citep{King62} equation.
\begin{equation}
\label{eq:1}
\rho(r)~=~\rho_{\rm bg}+\frac{\rho_{\rm o}}{1+\left(\sfrac{r}{r_{\rm c}}\right)^2},
\end{equation}
where, $r_{\rm c}$, $\rho_{\rm o}$, and $\rho_{\rm bg}$ are the core radius, central surface density, and background surface density respectively. We performed three radial density profiles (RDPs) as shown in Figure \ref{fig3} fitted with King’s model in Eq. \ref{eq:1}. All clusters' observable areas were separated into several concentric rings revolving around the cluster's core. The formula ($r_{\rm i}=N_{\rm i}/A_{\rm i}$) is used to calculate the number density ($r_{\rm i}$) in the i$^{\rm th}$ zone, where ($N_{\rm i}$) is the number of stars and ($A_{\rm i}$) is the area of the i$^{\rm th}$ zone.

In the following analysis, the limiting radius ($r_{\rm cl}$; arcmin) is defined as the distance from the cluster centre at which the gravitational pull from the cluster is equal to the gravitational attraction exerted by the centre of the Galaxy \citep{von1957}. $r_{\rm cl}$ is determined by comparing the cluster's stellar density profile, represented by the function $\rho(r)$, with the background density level, represented by the function $\rho_b$. The background density is defined as $\rho_b = \rho_{\rm bg} + 3\sigma_{\rm bg}$, where $\rho_{\rm bg}$ represents the background star density and $\sigma_{\rm bg}$ is the uncertainty in this background value, expressed in units of stars per arcmin$^{-2}$. This method provides a reliable measure of the cluster's boundary, at which the internal gravitational influence begins to diminish in comparison to the external Galactic forces. The limiting radius, denoted by $r_{\rm cl}$, is mathematically expressed by the following equation \citep{Bukowiecki11}:
\begin{equation}
\label{eq:2}
r_{\rm cl} = r_{\rm c}~\sqrt{\frac{\rho_{\rm o}}{3~\sigma_{\rm bg}} - 1}.
\end{equation}

Two gravitational forces act on any cluster to keep it bound: one is directed towards the Galactic center, and the other is directed towards the cluster center. The distance at which these two forces come into balance is known as the tidal radius ($r_{\rm t}$). It might thus serve as a separation between stars in a cluster that are gravitationally bound and those that are not. The effect of gravitationally large bodies in the Galactic disk, such as star clusters, was examined by \citet{Roser19}. The distance from the centre of a star cluster to the first Lagrangian point ($L_{\rm 1}$) \citep{Küpper08} known as the Jacobi radius ($r_{\rm J}$ (pc); Eq. \ref{eq:rJ}). It may therefore stay well within this radius, so that it does not indicate any obvious truncation in the stellar distribution, as it changes together with the cluster (growing, shrinking, moving to areas with strong or weak tides). For many clusters, it is possible to determine their Jacobi radii $r_{\rm J}$ more accurately than their tidal radii $r_t$. The main limitation is that the $r_{\rm J}$ fluctuates throughout a cluster's orbit; therefore, for extremely eccentric orbits, the present radius may not be a reliable indicator of the average $r_{\rm J}$ a cluster encounters, which establishes its mass-loss rate \citep{Holger10}.

\begin{equation}
\label{eq:rJ}
r_{\rm J}=\left[\frac{G~M_{\rm C}}{2~V_{\rm rot}^2}\right]^{1/3} \bigg(R_{\rm gc}\bigg)^{2/3}.
\end{equation}

where $G$ is the gravitational constant (= 4.30 $\times$ 10$^{-6}$ kpc M$^{-1}_{\odot}$ km$^2$ s$^{-2}$), $M_{\rm C}$ is the cluster mass (in M$_{\odot}$), ($V_{\rm rot} = 220$ km s$^{-1}$) is the rotational velocity of the Galaxy, and $R_{\rm gc}$ is the distance of the cluster from the Galactic centre.

Eq. \ref{eq:rJ} reflects our $r_J$ as 5.487 $\pm$ 2.45, 6.658 $\pm$ 2.58, and 10.444 $\pm$ 3.23 (pc) with respect manner of OCSN OCs and presented in Table \ref{tab2} into which we note our numerical values of $r_J$ for OCSN 203 and OCSN 213 are inline with those obtained by \cite{Hunt24} as listed in the same Table \ref{tab2}, where $r_{\rm J}$ of OCSN 244 is missed with \cite{Hunt24}.

For compact star clusters, \cite{Bonatto09} defined the density contrast parameter (i.e., $\delta_{\rm c}=1+\rho_{\rm o}/\rho_{\rm bg}$); which lags like ($7\le\delta_{\rm c}\le23$). Therefore, the estimated $\delta_{\rm c}$ are 10.60 $\pm$ 3.26 (OCSN 203), 2.86 $\pm$ 0.06 (OCSN 213), and 14.00 $\pm$ 3.74 (OCSN 244), and we may conclude that the OCSN 213 is a scattered cluster where its contrast parameter is less than its mean limiting values (i.e., $7\le\delta_{\rm c}\le23$). 

According to \cite{King66}, the concentration parameter (i.e., $C=r_{\rm cl}/r_{\rm c}$) is the ratio of the cluster limiting and core radii. It could be an indication of the cluster's center concentration. On the other hand, King's definition of the $C$ parameter from \cite{King66} was reversed by \cite{Santos12}. Young clusters are predicted to have low $C$ values according to \cite{Santos12}, since many of its members are still gathered in the center and have not had enough time to disperse to the edges. Our calculated $C$ range from 1.02 $\pm$ 0.01 to 4.74 $\pm$ 0.46 derived from Table \ref{tab2} is reasonable according to \cite{Santos12} and may concluded that the two first OCs are younger ones than third one. Additionally, according to \citep{Maciejewski07} research, $r_{cl}$ is ranged from 2 to 7 times the value of the $r_{\rm c}$, these values are inline with our estimation as shown in Table \ref{tab2} for OCSN 244.

\begin{table*}
\caption{New estimated values of the centre in equatorial and Galactic coordinates for the three OCSNs. The second portion represent the inner structural properties of the three OCSNs have been obtained with the most recent ones.}
\centering
\begin{tabular}{lccc}
\hline
{\bf Parameters}&{\bf OCSN 203}&{\bf OCSN 213}&{\bf OCSN 244}\\
\hline
\hline
$\alpha$ & 03$^h$44$^m$35$^s$.5 & ~05$^h$44$^m$ 35$^s$.5 & ~05$^h$35$^m$17$^s$.2  \\
$\delta$ & 32$^o$09'52''.2 & -07$^o$ 57'03''.6 & -05$^o$22'48''.0  \\
$l$  & 160$^o$.929& 212$^o$.784        & 209$^o$.269\\
$b$  & -17$^o$.257& -17$^o$.910        & -18$^o$.818   \\
\hline
$\rho_{\rm o}$ (stars arcmin$^{-2}$)   &	6.9133 $\pm$ 0.6729	& 3.7363 $\pm$ 0.6137   & 3.250 $\pm$ 0.0469 \\
$\rho_{\rm bg}$ (stars arcmin$^{-2}$)&	0.7201 $\pm$ 0.6739	& 2.0100 $\pm$ 0.6137	&  0.2501 $\pm$ 0.0469 \\
r$_{\rm c}$ (arcmin) &	2.28$^{+0.98}_{-0.89}$ & 10.31$^{+1.96}_{-2.10}$		&  1.68$^{+0.17}_{-0.17}$	 \\
r$_{\rm c}$ (pc) &	0.210 & 1.284 & 0.190 \\
r$_{\rm c}$ (pc)$^*$  &	0.36&3.85&-- \\
 r$_{\rm cl}$ (arcmin)&	3.55 &  10.46 &  7.90 \\
 r$_{\rm cl}$ (pc)&	0.327 &  1.302 &  0.900 \\
 r$_{\rm t}$ (arcmin)&	5.06$^{+1.10}_{-0.96}$	& 13.52$^{+1.04}_{-1.07}$		&  14.16$^{+0.92}_{-0.90}$ \\
r$_{\rm t}$ (pc)&	0.467 & 1.683 & 1.600 \\
 r$_{\rm J}$ (pc)&	5.985 $\pm$ 2.45 & 6.658 $\pm$ 2.58 & 10.444 $\pm$ 3.23 \\
 r$_{\rm J}$ (pc)$^{*}$&	8.087 & 5.288 & -- \\
$\delta_{\rm c}$& 10.60 $\pm$ 3.26	& 2.86 $\pm$ 0.06 &	14.00 $\pm$ 3.74	\\
$C$& 1.56 $\pm$ 0.08	& 1.02 $\pm$ 0.01 &	4.74 $\pm$ 0.46	 \\  
\hline
$^*$\cite{Hunt24} & & & \\
\end{tabular}
\label{tab2}
\end{table*}

\subsection{Astrometric parameters and distance estimation}
\label{sec:astro_parmeters}
The distance and age of the OCs are the most important factors in tracing the Galactic structure and the chemical growth \citep{1993A&A...267...75F}. The clusters' CMDs have proven to be a useful tool for locating the member stars and determining the basic cluster properties (e.g., reddening, distance modulus, and ages). In this study, we employed the {\sc UPMASK} method to analyze the five-dimensional astrometric space of stars within the OCSN clusters, encompassing equatorial coordinates, proper motions, and parallaxes. This approach allowed us to accurately assess the spatial and kinematic properties of the cluster members, providing a robust framework for distinguishing between true members and field stars. The $k$-means clustering value was set between 10 and 30 in order to take account of different star densities, and 1000 iterations were run to ensure robust results \citep{YontanCanbay2023}. Membership probabilities were computed and visualised via histograms, in accordance with the methods of \citet{Cantat2018}. The {\sc UPMASK} approach improves the accuracy of the membership determination. For this study, only stars with membership probabilities ($P \ge 50\%$) are assigned as cluster most probable candidates. Therefore, we have 227, 200, and 551 with OCSN 203, 213, and 244 respectively. \cite{Hunt24} deduced that membership is 289 (OCSN 203) and 71 (OCSN 213). Under or overestimated  member stars retrieved between us and \cite{Hunt24} are due to the method of estimation.

\begin{figure*}
\begin{center}
\includegraphics[width=0.85\linewidth]{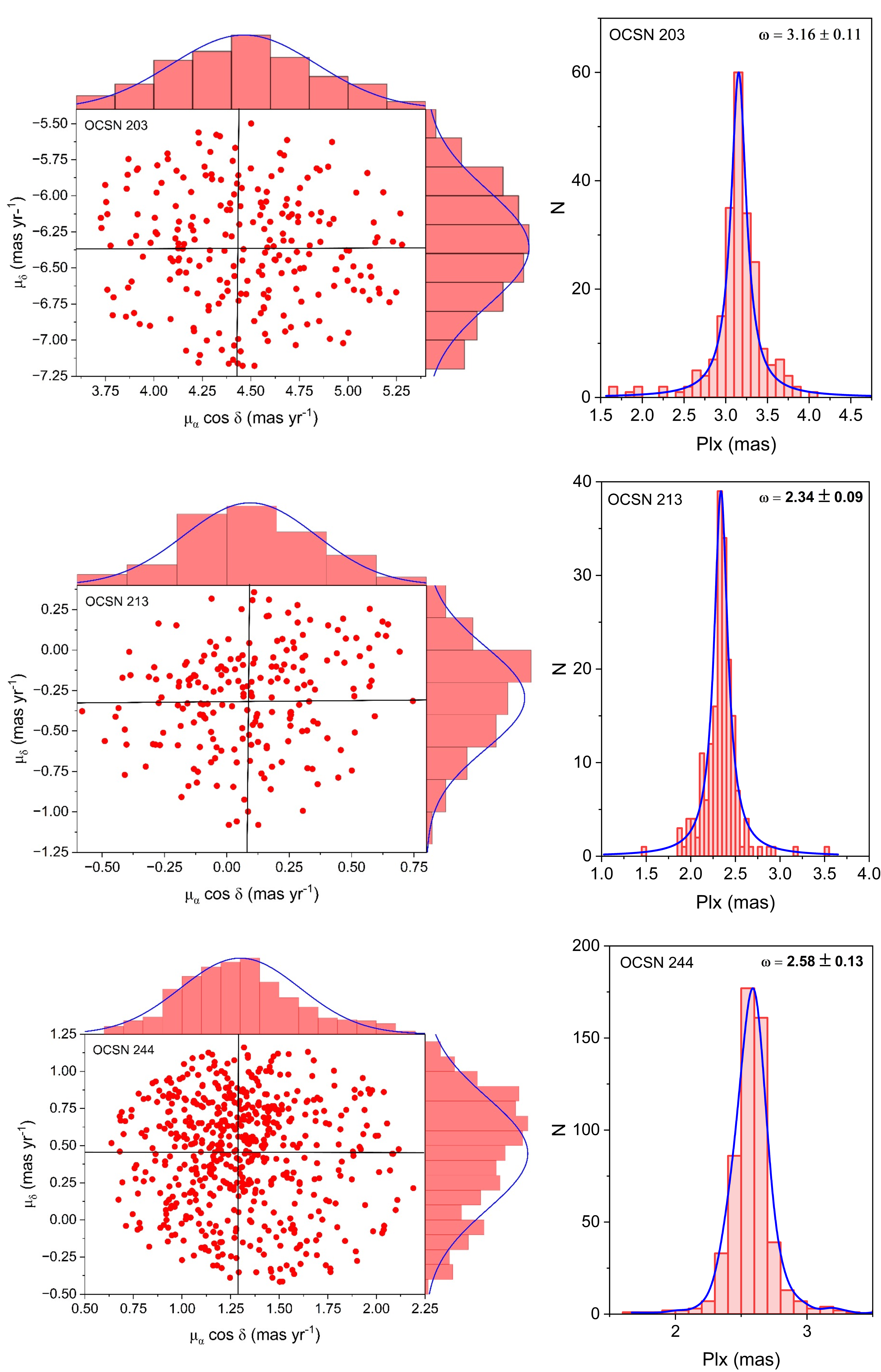}
\caption {Mean proper motion distribution and their concerned Gaussian fitting of parallax (mas) for OCSN 203 (upper panel), OCSN 213 (middle panel), and OCSN 244 (lower panel).}
\label{fig4}
\end{center}
\end{figure*}

In what follows with adopted members, we determined the mean proper motion on both sides ($\mu_\alpha\cos\delta,~\mu_\delta$; mas yr$^{-1}$) by stellar space distribution like drawn in the vertical left panel and their contours as a middle panel of Figure \ref{fig4} and it numerically gives (4.46, -6.36; OCSN 203), 0.09, -0.29; OCSN 213), and (1.30, 0.44; OCSN 244) in mas yr$^{-1}$ units. The vertical right panel is concerned with those Gaussian distributions as histograms like (3.16 $\pm$ 0.11; OCSN 203), (2.34 $\pm$ 0.09; OCSN 213), and (2.58 $\pm$ 0.13; OCSN 244) in mas units, which reflect the astrometric distance as 317 $\pm$ 18, 428 $\pm$ 21, and 388 $\pm$ 20 pc for OCSN 203, OCS 213, and OCSN 244, respectively.

\section{Reddening, Distance Modulus and Age}
\label{sec4}
Accurately identifying cluster star members hinges on the careful fusion of radial velocities, distances, and proper motions, either as standalone markers or through their harmonious interplay. The sweeping reach of large spectroscopic surveys, designed with diverse aims in mind, now offers the tools to uncover these members with new precision \citep{Allende08,Gilmore12,De_Silva15}. Yet, spectroscopic observations demand telescope time—a rare and coveted currency—limiting our ability to peer into the deeper nature of many star clusters. Moreover, no comprehensive survey is on the horizon to gather radial velocities for all the stars in the Milky Way. Still, a constellation of methods exists to separate the stars birthed in the same cluster from the surrounding field stars, as these stellar siblings carry traces of their common origin in space (e.g., \cite{Krone-Martins14,Javakhishvili06,Balaguer98}). Most of these approaches center around analyzing proper motions, weaving the stars' subtle movements into a map of shared trajectories.

In their study, \cite{Perren15} employed the {\sc ASteCA} code to estimate the total number of probable cluster members using two distinct methods. The first method involves calculating the integral of the radial density profile (RDP) from zero to the $r_t$, adjusted for the estimated star field density. This approach relies on fitting a three-parameter (3P) King profile, which provides a model for the density distribution of stars in the cluster. For this method to yield reliable results, it is essential to have a reasonable estimate of the tidal radius and for the 3P fit to converge adequately; otherwise, there is a significant risk of overestimating the number of cluster members. The second method is more straightforward and relies on counting stars ($n_{\rm fl}$) that are likely field stars within the cluster region. This count is calculated by multiplying the field star density ($d_{\rm field}$) by the area ($A_{\rm cl}$) of the cluster, which is determined using the $r_{\rm cl}$. The final estimated number of cluster members, $n_{\rm cl}$, is obtained by subtracting this amount from the actual number of stars within the $r_{\rm cl}$ boundary ($n_{\rm cl+fl}$):

\begin{equation}
n_{\rm cl}=c_{\rm cl+fl}-d_{\rm field}A_{\rm cl}.
\end{equation}

\begin{figure*}
\begin{center}
\includegraphics[width=0.75\linewidth]{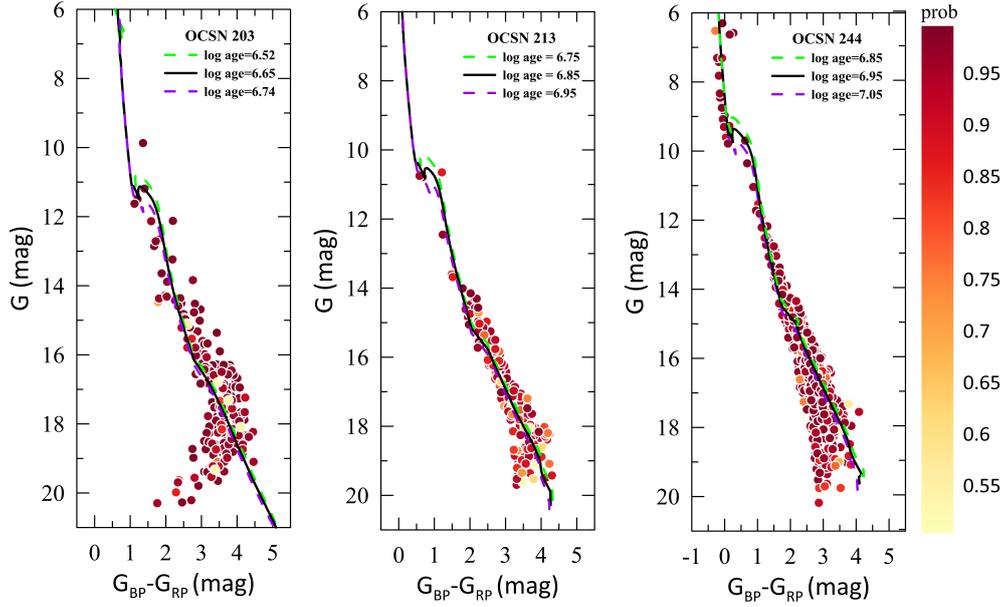}
\caption {The CMDs of OCSN 203 (left panel), OCSN 213 (middle panel), and OCSN 244 (right panel). The fitted extinction was corrected using \cite{Bressan12} methodology.}
\label{fig5}
\end{center}
\end{figure*}

Both methodologies for estimating cluster membership are influenced by the completeness of the data, as they provide approximate counts of members based on the lowest observed magnitudes. In this study, we utilized the $Gaia$ DR3 photometric magnitudes ($G,~G_{\rm BP},~G_{\rm RP}$) for our candidate stars. For each CMD of the clusters, we employed the {\sc ASteCA} code alongside the PARSEC v1.25 theoretical isochrones \citep{Bressan12} to derive the cluster metallicity ($Z$) and ages on a logarithmic scale. The best-fit metallicities obtained for the clusters were 0.01308 $\pm$ 0.00564, 0.01394 $\pm$ 0.00054, and 0.01413 $\pm$ 0.00607 for OCSN 203, OCSN 213, and OCSN 244, respectively. Correspondingly, the log ages (in years) were determined to be 6.65 (with a range of 6.52 to 6.74), 6.85 $\pm$ 0.10, and 6.95 $\pm$ 0.10 for OCSN 203, OCSN 213, and OCSN 244. Figure \ref{fig5} presents the fitted CMDs for the color-magnitude pair ($G_{\rm BP} - G_{\rm RP}, G$) magnitudes. By leveraging the high-probability members identified from $Gaia$ DR3, we estimated the reddening values using the magnitudes $G_{\rm BP}$ and $G_{\rm RP}$. The relationship used for estimating the reddening was $E(G_{\rm BP} - G_{\rm RP}$) = 1.289 $\times$ $E(B - V)$ \citep{2018MNRAS.479L.102C,2019A&A...624A..34Z}. In our analysis, the observed data were corrected for reddening using a line-of-sight extinction coefficient of $A_{\rm G} = 2.74 \times E(B-V)$. The calculated reddening values were found to be $E(G_{\rm BP}-G_{\rm RP}$) = 0.2836 $\pm$ 0.06, 0.5672 $\pm$ 0.07, and $0.7476 \pm 0.08$ for OCSN 203, OCSN 213, and OCSN 244, respectively.

One of the main justifications for our CMDs is that they show the distance moduli ($m-M$) are 7.603 $\pm$ 0.363, 8.615 $\pm$ 0.341, and 8.522 $\pm$ 0.343 mag for OCSN 203, OCSN 213, and OCSN 244 respectively, thus reflecting the photometric distances ($d_{\rm phot};$ pc) by about 332 $\pm$ 18, 529 $\pm$ 23, and 506 $\pm$ 23 for OCSN 203, OCSN 213, and OCSN 244 respectively, which is inline with those obtained by \cite{Qin23}. Table \ref{tab3} displays our obtained astrophysical parameters.

\begin{table*}
\caption{The astrophysical and photometric parameters of OCSN 203, OCSN 213, and OCS 244 were obtained and compared with the findings presented in \cite{Qin23} and \cite{Hunt24}.}
\centering
\small
\begin{tabular}{lcccl}
\hline
{\bf Parameters}&{\bf OCSN 203}&{\bf OCSN 213}&{\bf OCSN 244}&{\bf Ref.}\\
\hline
\hline
N&227	&200	&551	&  This study\\[0.8 ex]
&289	&71&	--	&\cite{Hunt24}\\[0.8 ex]
$\mu_{\alpha}\cos{\delta}$ (mas yr$^{-1}$)& 4.46 $\pm$ 0.06	&0.09 $\pm$ 0.04	&1.30 $\pm$	0.07 &  This study\\[0.8 ex]
&4.47 $\pm$ 0.55&	0.10 $\pm$ 0.27	&1.41  $\pm$ 0.78	&\cite{Qin23}\\[0.8 ex]
&4.423  $\pm$ 0.026	&0.074 $\pm$ 0.023	&--  	&\cite{Hunt24}\\[0.8 ex]
$\mu_{\delta}$ (mas yr$^{-1}$)&-6.36 $\pm$ 0.03	&-0.29	$\pm$ \textbf{0.04}&0.44	$\pm$ 0.06 &  This study\\[0.8 ex]
&-6.30  $\pm$ 0.57	&-0.33  $\pm$ 0.32	&0.35  $\pm$ 0.89	&\cite{Qin23}\\[0.8 ex]
&-6.343  $\pm$ 0.030&	-0.328  $\pm$ 0.030	&--	&\cite{Hunt24}\\[0.8 ex]
$\varpi$ (mas)&3.16 $\pm$ 0.11	& 2.34 $\pm$ 0.09	& 2.58 $\pm$ 0.13	&  This study\\[0.8 ex]
&3.17  $\pm$ 0.13	&2.35  $\pm$ 0.07	&2.49  $\pm$ 0.13&	\cite{Qin23}\\[0.8 ex]
&3.164  $\pm$ 0.006&	2.354  $\pm$ 0.005	&--	&\cite{Hunt24}\\[0.8 ex]

$d_{\rm \varpi}$ (pc)& 317  $\pm$ 18	& 428  $\pm$ 21	& 388  $\pm$ 20&	  This study\\[0.8 ex]

&316  $\pm$ 0.03	&426  $\pm$ 0.03&	402  $\pm$ 0.05&	\cite{Qin23}\\[0.8 ex]
&316.06	&424.81	&--	&\cite{Hunt24}\\[0.8 ex]

$Z$&0.01308  $\pm$ 0.00564	&0.01394  $\pm$ 0.00054	&0.01413  $\pm$ 0.00607	&  This study\\[0.8 ex]

$\log t$ (yr)& 6.65$^{+0.09}_{-0.13}$	& 6.85  $\pm$ 0.10	& 6.95  $\pm$ 0.10	&  This study\\[0.8 ex]

&6.25	&6.70	&6.65	&\cite{Qin23}\\[0.8 ex]
&6.940	&7.306	&--	&\cite{Hunt24}\\[0.8 ex]

$E(B-V)$ (mag)&0.22  $\pm$ 0.06&	0.44  $\pm$ 0.06	&0.58  $\pm$ 0.08	&  This study\\[0.8 ex]

&0.11	&0.48&	0.53	&\cite{Qin23}\\[0.8 ex]
&0.814	&0.770&	--	&\cite{Hunt24}\\[0.8 ex]

$E(G_{\rm BP}-G_{\rm RP})$ (mag)&0.2836  $\pm$ 0.06&	0.5672  $\pm$ 0.07	&0.7476  $\pm$ 0.08	&  This study\\[0.8 ex]

$(m-M)$&7.603  $\pm$ 0.363&	8.615  $\pm$ 0.341&	8.522  $\pm$ 0.343	&  This study\\[0.8 ex]
&7.79	&9.46&	9.39	&\cite{Qin23}\\[0.8 ex]
&7.60	&8.22	&--	&\cite{Hunt24}\\[0.8 ex]
$d_{\rm photo}$ (pc)&332  $\pm$ 18	&529  $\pm$ 23	&506  $\pm$ 23&	  This study\\[0.8 ex]
&313.238&418.382&--&\cite{Hunt24}\\[0.8 ex]

$X_{\odot}$ (kpc)& -0.286  $\pm$ 0.017&	-0.342  $\pm$ 0.018 &	-0.320  $\pm$ 0.018&	  This study\\[0.8 ex]

&-0.281	&-0.334	&--	&\cite{Hunt24}\\[0.8 ex]

$Y_{\odot}$ (kpc)& 0.099  $\pm$ 0.010&	-0.221  $\pm$ 0.015	& -0.180  $\pm$ 0.013	&  This study\\[0.8 ex]

&0.010	&-0.212	&--	&\cite{Hunt24}\\[0.8 ex]

$Z_{\odot}$ (kpc)& -0.094  $\pm$ 0.010 &	-0.132  $\pm$ 0.012&	-0.125  $\pm$ 0.011	&  This study\\[0.8 ex]

&-0.096&	-0.136&	--	&\cite{Hunt24}\\[0.8 ex]

$R_{\rm gc}$ (kpc)& 8.487  $\pm$ 0.092	& 8.545  $\pm$ 0.092&	8.523  $\pm$ 0.003	&  This study\\[0.8 ex]    
\hline
\label{tab3}
\end{tabular}
\end{table*}

Taking into account our estimated distances $d_{\rm plx}$, we infer to include those distances to the Galactic center $R_{\rm gc}$ (i.e., $R_{\rm gc}=\sqrt{R_{\rm o}^2+(d\cos{b})^2-2R_od\cos{b}\cos{l}}$) where $R_{\rm o}$ = 8.20 $\pm$ 0.10 kpc \citep{Bland19}. Using this approach, we can derive the projected distances towards the Galactic plane, denoted as ($X_{\rm \odot},~Y_{\rm \odot}$) as well as the distance from the Galactic plane ($Z_{\rm \odot}$). The calculations are performed according to specific relationships, and the results obtained are summarized in Table \ref{tab4}.
\begin{equation}
\label{eq:Xsun}
\begin{split}
X_{\odot} &= d~cosb~cosl, \\
Y_{\odot} &= d~cosb~sinl, \\
Z_{\odot} &= d~sinb. 
\end{split}
\end{equation}

\section{Luminosity and mass functions}\label{sec5}
\begin{figure}
\begin{center}
\includegraphics[width=1\linewidth]{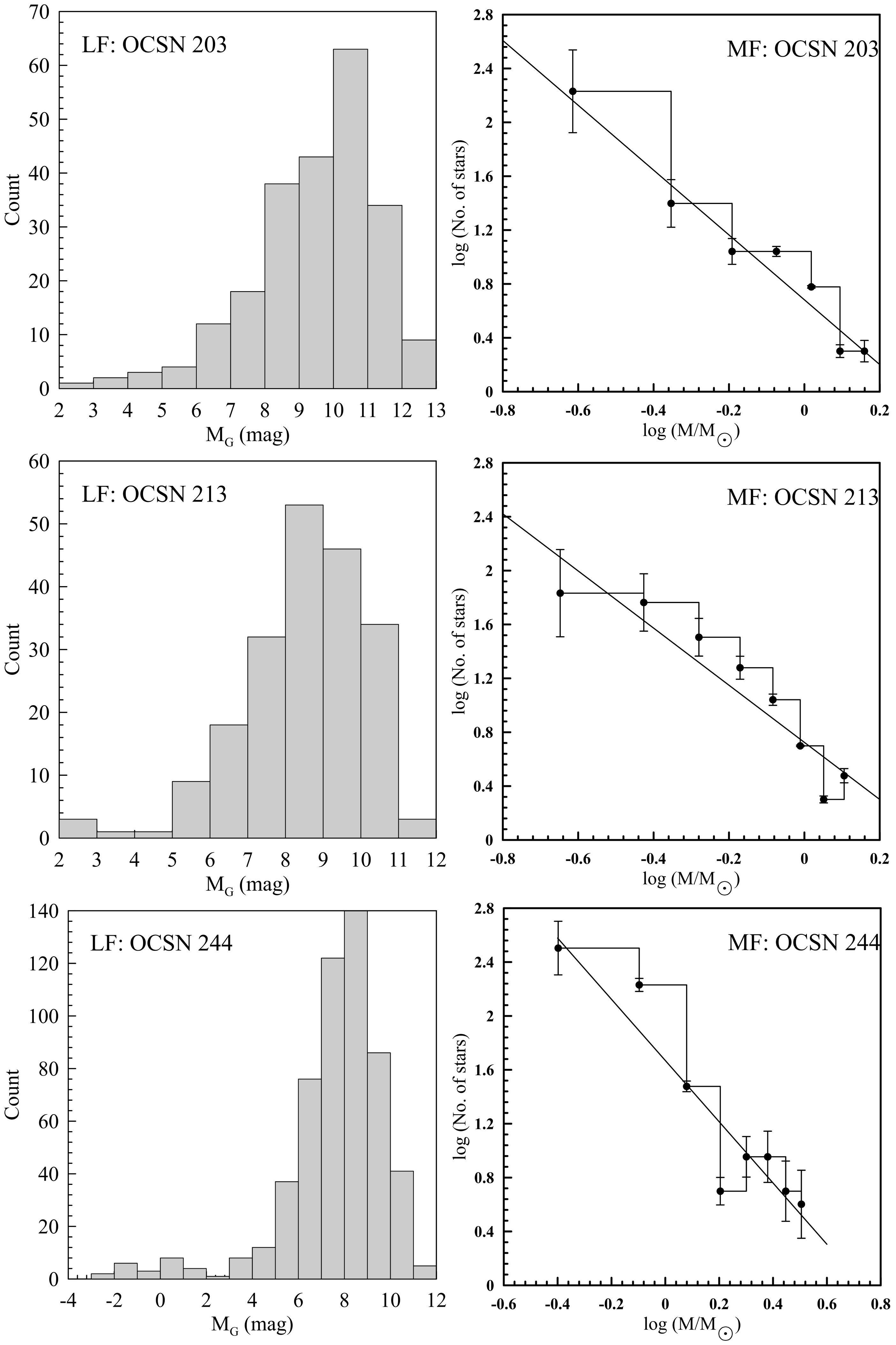}
\caption {The vertical left panel presents the true LFs for OCSN’s clusters, which were built by converting the observed $G$ magnitudes of member stars into the absolute magnitudes ${M_{\rm G}}$ considering the distance modulus. ($m-M$) of the cluster. The vertical right panel shows MFs derived using the most probable members where the solid lines indicate the fitting of the power-law given by \cite{Salpeter55}.}
\label{fig6}
\end{center}
\end{figure}

Based on recent worksheet row data of DR3 \citep{2021A&A...649A...1G} with OCSN’s aforementioned. We have obtained new center positions as well as the physical and photometric parameters. We can infer to estimate the luminosity function (LF) and mass function (MF). The LF is defined as the distribution of member stars according to different absolute magnitudes, as the members of each cluster are formed under similar physical conditions (same morphology) from the same molecular cloud at the same time; Figure \ref{fig6} with vertical left panel displays the LF of those member stars. From the LF of each cluster, we can estimate the whole mean cluster's absolute magnitudes ($\overline{M_{\rm G}}$; mag) as 9.54 $\pm$ 3.09 (OCSN 203), 8.52 $\pm$ 2.92 (OCSN 213), and 7.60 $\pm$ 2.76 (OCSN 244). 

The mass-luminosity relation (MLR) is a well-established correlation between the absolute magnitudes ($M_{\rm G}$; mag) and collective masses ($M_{\rm C}; M_{\odot}$) of astronomical objects. By accounting for absolute magnitudes $M_{\rm G}$ and the collective masses $M_C$ devoted with adopted isochrones \citep{Evans18} on CMDs for estimated ages, distance modulus, and reddening, we can infer the MLR of individual member stars with the second-order polynomial function in the following form.
\begin{equation}
M_{\rm C}=a_{\rm 0}+a_{\rm 1}M_{\rm G} +a_{\rm 2}M_{\rm G}^2.                   
\end{equation} 

Here the constants ($a_{\rm 0},~a_{\rm 1},~a_{\rm 2}$) and their uncertainties as well as the mass parameters (i.e., total masses $M_{\rm C}$ and the mean masses $\langle{M_{\rm C}}\rangle$ are given in Table \ref{tab4} in solar units.

Mass spectrum of OCs contains very low and/or high mass stars, this makes them the ideal objects to study initial mass function (IMF), which is defined as the primary arrangement of the star’s masses \citep{Scalo98,Phelps93,Durgapal01,Piatti02,Piskunov04,Sung04,Yadav02,Yadav04,Bisht17,Bisht19}. In their seminal work, \cite{Salpeter55} defined the IMF as the total number of stars ($dN$) distributed across a logarithmic mass scale in a given mass bin ($dM$) with a central mass ($M$), where the number of stars per unit mass is given by the power law $dN/dM=M^{-\alpha}$ with an exponent of 2.35. The present-day mass function MF can be expressed mathematically as:
\begin{equation}
\label{eq:mf}
log~\bigg(\frac{dN}{dM_{\rm G}}\bigg)~=-\alpha~log(M_{\rm G}) ~ + ~\text{constant}
\end{equation}
where $\alpha$ is a dimensionless quantity representing the slope of the straight line that describes the MF (shown in the vertical right panel of Figure \ref{fig6}), is of great significance for comprehending the mass distribution of stars. \cite{Salpeter55} initially determined $\alpha = 1.35$, which is now recognized as a signature of dynamical evolution for massive stars ($\geq 1 M_{\odot}$). In accordance with Salpeter's power law,  the number of stars within each mass range decreases rapidly as mass increases. Our calculated slopes ($2.13 \leq \alpha \leq 2.41$), obtained through least-square fitting of MF data, are consistent with the findings of \cite{Salpeter55}.

Based on our analysis, the absolute magnitudes with our OCs range from 2.269 to 12.692 mag (OCSN 203), from 2.033 to 11.641 mag (OCSN 213), and from -2.220 to 11.658 mag (OCSN 244). According to our estimation of MF, we concluded that the masses range like  ($0.146 \ge M_{\rm C} \ge 2.140$; OCSN 203), ($0.031\ge M_{\rm C} \ge1.875$; OCSN 213), and ($0.037 \ge M_{\rm C} \ge 3.048$; OCSN 244).

\begin{table}
\caption{The estimated absolute magnitudes from luminosity functions (LF), masses with MLR fitting, and slopes of mass functions (MF) for OCSN clusters are presented herewith.}
\centering
\tiny
\begin{tabular}{lccc}
\hline
{\bf Parameters}&{\bf OCSN 203}&{\bf OCSN 213}&{\bf OCSN 244}\\   
\hline
\hline
$\overline{M_{\rm G}}$ (mag)&9.54 $\pm$ 3.09 &	8.52  $\pm$ 2.92 &	7.60  $\pm$ 2.76 \\
$a_0$& 3.3007 $\pm$ 0.006 & -0.5700 $\pm$ 0.001  & 0.0257 $\pm$ 0.001   \\
$a_1$& 2.504 $\pm$ 0.007 & -0.3300 $\pm$ 0.001  & 0.0100 $\pm$ 0.001   \\
$a_2$& 2.360 $\pm$ 0.004 & -0.2923 $\pm$ 0.002  & 0.0080 $\pm$ 0.001   \\
$M_{\rm C}$ ($M_{\odot}$)& 67  $\pm$ 8.19 &	91  $\pm$ 9.54 &	353  $\pm$ 18.79 \\
$M_{\rm C}$ $^*$ &191.70  $\pm$ 38.10   &	51.19  $\pm$ 16.99 & 	-- \\
$\langle{M_{\rm C}}\rangle$ ($M_{\odot}$)&0.290	& 0.452 &	0.641 \\
$\alpha$&2.41  $\pm$ 0.06 &	 2.13  $\pm$ 0.07 &	2.28  $\pm$ 0.07\\
\hline
\multicolumn{4}{l}{$^*$\cite{Hunt24}} \\   
\end{tabular}
\label{tab4}
\end{table}

\section{OCSN’s Dynamics and its Kinematics}
\label{sec6}
\subsection{Dynamical Evolution Times}
Dynamically, the interaction between stars within OCs results in energy exchange, as described by \cite{Inagaki85} and \cite{Baumgardt03}, which is different from the behavior seen in more compact systems like globular clusters. Open clusters are characterized by a looser spatial distribution, and as a result, forces of contraction or destruction lead to the mass segregation phenomenon, where more massive stars tend to move toward the cluster core, while fainter stars are found farther from the center. This effect has been observed in numerous open clusters in recent studies \citep{Piatti19,Zeidler17,Dib18,Rangwal19,Bisht20,Joshi20}. As this process unfolds, the velocity distribution of the cluster approaches a Maxwellian equilibrium, contributing to the system's dynamical stability \citep{Yadav13,Bisht19}. The timescale required for this dynamical evolution, referred to as the dynamical relaxation time ($T_{\rm relax}$; Myr), is mathematically defined by \cite{Spitzer71}. It depends on both the number of member stars ($N$) and core $r_c$ $\&$ tidal $r_t$ radii \citep{Spitzer71}.
\begin{equation}
\label{eq:Trelax}
T_{\rm relax} ~=~ \frac{8.9\times10^5~N^{1/2}~ R_{\rm h}^{3/2}}{\langle{M_{\rm C}}\rangle^{1/2}~log(0.4~N)}
\end{equation}
where $R_{\rm h}$ is the radius (in pc) containing $\sim50\%$ of the cluster mass and can be estimated based on the transformation given in \cite{
2006BaltA..15..547S},
\begin{equation}
R_{\rm h}~=~0.547 \times r_{\rm c} \times \Big(\frac{r_{\rm t}}{r_{\rm c}}\Big)\textsuperscript{0.486}.
\end{equation}

Where $r_{\rm c}$ and $r_{\rm t}$ represent the core and tidal radii, respectively. Based on these parameters, the computed half-mass radii ($R_{\rm h}$) for OCSN 203, OCSN 213, and OCSN 244 are 0.170, 0.801, and 0.292 pc, respectively. Consequently, the corresponding dynamical relaxation times ($T_{\rm relax}$) are 0.884, 7.051, and 1.286 Myr for these clusters. The dynamical state of a cluster can be quantified by its dynamical evolution parameter ($\tau = \text{age} / T_{\rm relax}$), when its much greater than one (i.e., $\tau \gg 1$) the cluster well-known as relaxed cluster and vice versa. In our consideration, it's clear that OCSN 203 and OCSN 244 are relaxed OCs while OCSN 213 are non-relaxed ones (i.e., $\tau = 1$).

In what follows, we focus on estimating the evaporation time ($\tau_{\rm ev} \approx 10^2 \times T_{\rm relax}$; Myr), which defines the time required for internal stellar encounters to eject all member stars from the cluster \citep{Adams01}. Low-mass stars typically escape from the cluster at low velocities, primarily through Lagrange points \citep{Küpper08}. To maintain its bound state, the cluster must have an escape velocity ($V_{\rm esc}$; km s$^{-1}$), which is determined by \bigg($V_{\rm esc} = R_{\rm gc}~ \sqrt{2~ G~ M_{\rm C} / 3 r_{\rm t}^3}$\bigg) \citep{Fich91,Fukushige00}, where $G = 4.30 \times 10^{-6}$ kpc M$^{-1}_{\odot}$ km$^2$ s$^{-2}$ is the gravitational constant. The computed results for the dynamical parameters, timescales, and escape velocities are listed in Table \ref{tab5}.

\begin{table}
\caption{Obtained different OCSN’s dynamical evolving times and their escaping velocities.}
\centering
\small
\begin{tabular}{llll}
\hline
{\bf Parameters}&{\bf OCSN 203}&{\bf OCSN 213}&{\bf OCSN 244}\\   
\hline
\hline
$T_{relax}$~(Myr)& 0.884 &	7.051  &	1.286 \\[0.8 ex]
$\tau$ & 5.06 &	1.00 &	6.93 \\[0.8 ex]
$\tau_{ev}$~ (Myr) & 88.40   &	705.13 & 128.58 \\[0.8 ex]
$V_{esc}$ (km s$^{-1}$) & 11671$\pm$108	& 1999$\pm$45 &	5907$\pm$77 \\[0.8 ex]
\hline
\label{tab5}
\end{tabular}
\end{table}

\subsection{CP}
\label{CP}
Kinematically, virtual star members appear as coherent and associated with moving OCs sharing similar properties, such as distance, kinematics, chemical composition, and age, as well as the radial velocity ($V_{\rm r}$; km s$^{-1}$). The determination of the point ($A_{\rm o},~D_{\rm o}$) by which the action of the stars in the cluster will merge (i.e., apex) is one of the main important tasks in the kinematical and physical examination \citep{Wayman65,Eggen84,Gunn88}. Several techniques are available for this purpose (including, i) the classical convergent point method \citep{Boss08}, ii) the AD chart method \citep{Chupina01,Chupina06}, and iii) the convergent point search method (CPSM) \citep{Galli12}. In this study, we specifically focus on the AD chart method (also known as the stellar apex method), which relies on the distribution of individual apexes of member stars within the equatorial coordinate system. This procedure was known and developed \citep{Chupina01,Chupina06} and has been frequently used \citep{Vereshchagin14,Elsanhoury18,Postnikova20,Elsanhoury22} in the direction of determining apexes and other numerous kinematical parameters, as well as their central kinematical substructures (e.g., M67, NGC 188, Pleiades, Castro $\&$ UMa, and IC 2391). In this manner,
\begin{equation}
\label{eq:ad}
\begin{split}
A_o &= \tan^{-1} \Bigg(\frac{\overline V_{\rm y}}{\overline V_{\rm x}}\Bigg) \\
D_o &= \tan^{-1} \Bigg(\frac{\overline V_{\rm z}}{\sqrt {\overline V_{\rm x}^2 + \overline V_{\rm y}^{2}}}\Bigg)
\end{split}
\end{equation}
where ($V_{\rm x},~V_{\rm y},~V_{\rm z}$; km s$^{-1}$) are stellar velocities along the ($x,~y,~z$) axes concerning the Sun. These velocities include the proper motions along proper motions in both directions and radial velocities whose estimated by \cite{Soubiran18} (see $\S\ref{sec:dynnamic}$).
\begin{equation}
\label{eq:Vs}
\begin{split}
V_{\rm x} & = -4.74~d~\mu_\alpha\cos\delta \sin\alpha~-~4.74~d~\mu_\delta\sin\delta \cos\alpha \\
&~+~V_{\rm r} \cos\delta \cos\alpha, \\ 
V_{\rm y} & = +4.74~d~\mu_\alpha\cos\delta \cos\alpha~-~4.74~d~\mu_\delta\sin\delta \sin\alpha \\
&~+V_{\rm r} \cos\delta \sin\alpha, \\
V_{\rm z} & = +4.74~d~\mu_\delta\cos\delta~+~V_{\rm r}\sin\delta.
\end{split}
\end{equation}

\begin{figure*}
\begin{center}
\includegraphics[width=0.85\linewidth]{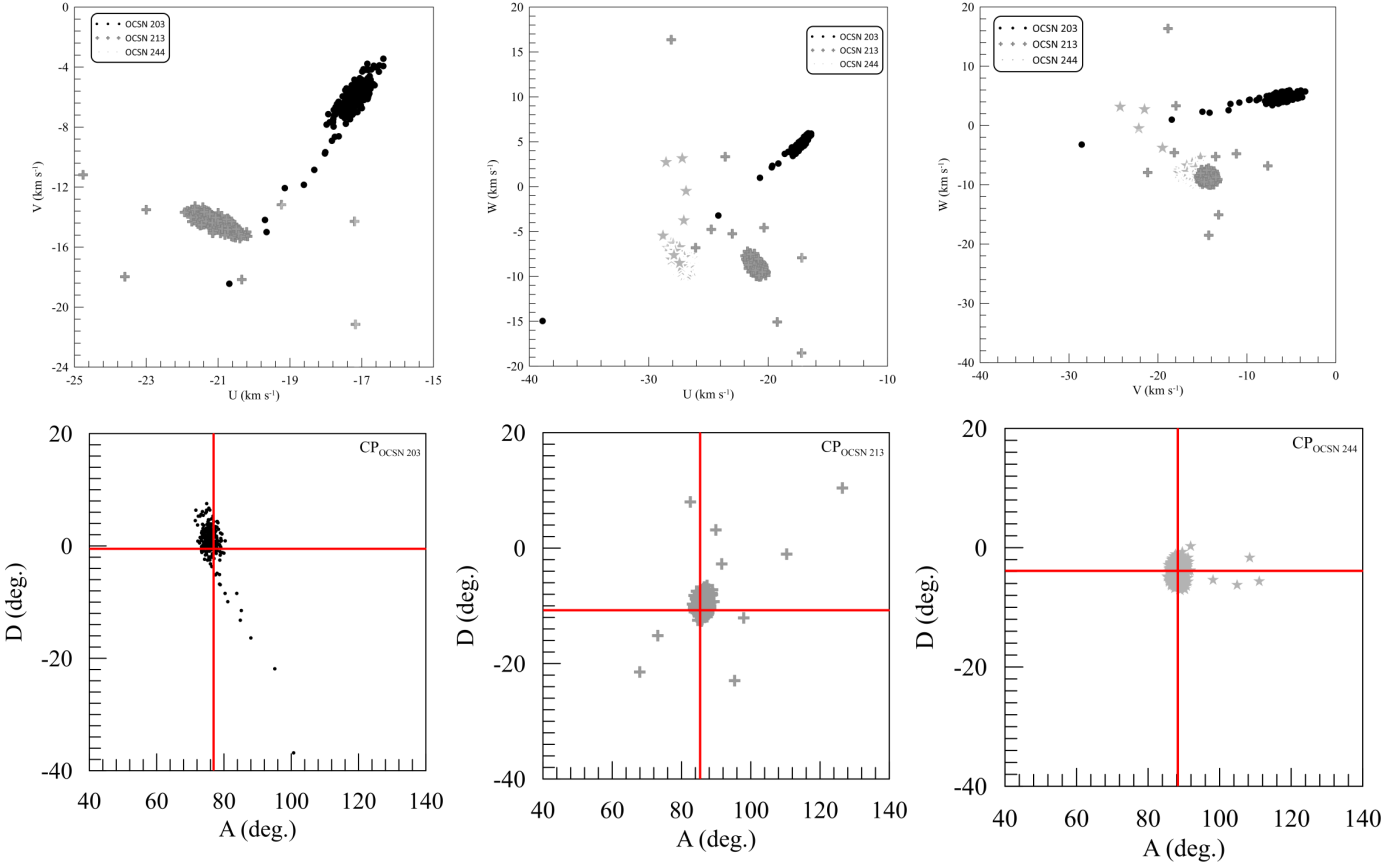}
\caption {The \textbf{top} panel shows the spatial velocity components ($U,~V,~W$) in Galactic coordinates for the cluster candidates. The \textbf{left} panel displays AD diagrams, with a cross marking the convergent point ($A_{\rm o},~D_{\rm o}$); see figure key.}
\label{fig7}
\end{center}
\end{figure*}

To derive the components of space velocities ($U,~V,~W$; km s${^{-1}}$) in Galactic coordinates, we utilized an equatorial-to-Galactic transformation matrix based on the SPECFIND v2.0 catalogue of radio continuum spectra. The space velocity components, calculated using Eqs. \ref{eq:Vs}, are transformed into Galactic coordinates through this matrix. The transformation follows the equations provided in \cite{Liu11}, specifically Eqs. (14):
\begin{equation}
\label{eq:uvw}
\begin{split}
U &=-0.0518807421 V_{\rm x}-0.8722226427 V_{\rm y} \\
&-0.4863497200 V_{\rm z},\\ 
V &=+0.4846922369 V_{\rm x}-0.4477920852 V_{\rm y} \\
&+0.7513692061 V_{\rm z},\\ 
W &=-0.8731447899 V_{\rm x}-0.1967483417 V_{\rm y} \\
&+0.4459913295 V_{\rm z},
\end{split}
\end{equation}
The upper panel of Figure \ref{fig7} presents the distribution of the space velocities along Galactic coordinates, numerically the mean values are ($\overline{U},~\overline{V},~\overline{W}$) = (-17.37 $\pm$ 4.17, -6.63 $\pm$ 2.57, -7.95 $\pm$ 2.82 km s$^{-1}$), (-21.03 $\pm$ 4.59, -14.30 $\pm$ 3.78, -8.96 $\pm$ 2.99 km s$^{-1}$), and (-27.33 $\pm$ 5.23, -15.81 $\pm$ 3.98, -8.33 $\pm$ 2.89 km s$^{-1}$) for OCSN's with respective manner. The lower panel shows the apex position ($A_{\rm o},~D_{\rm o}$) of these clusters with an AD chart method with comments as well as the distribution of the space velocities for these clusters. Our numerical values of the convergent point are (76$^o$.77 $\pm$ 0$^o$.01, -0$^o$.23 $\pm$ 0$^o$.00; OCSN 203), (85$^o$.71 $\pm$ 0$^o$.11, -9$^o$.63 $\pm$ 0$^o$.03; OCSN 213), and (88$^o$.19 $\pm$ 0$^o$.11, -4$^o$.04 $\pm$ 0$^o$.01; OCSN 244).

\subsection{Dynamic Orbit Parameters and Galactic Population}
\label{sec:dynnamic}

In our dynamic orbit analysis, we used the {\sc MWPotential2014} model integrated in the well-known {\sc galpy}\footnote{https://galpy.readthedocs.io/en/v1.5.0/} package. This Python-based tool is widely recognised for its efficiency in modelling Galactic dynamics as developed and described by \citet{Bovy_2015}. The model includes several significant parameters to accurately simulate the gravitational potential of the MW. The methodology applied in this study, which has also been validated through prior studies of individual stars and other star clusters, enables an extensive examination of the dynamical properties and spatial distribution of OCs within our Galaxy \citep{Tasdemir_2023, Yucel_2024}. These include the Galactocentric distance of the Sun ($R_{\rm o} = 8.20~\pm~0.10$ kpc) and its rotational velocity ($V_{\rm rot} = 220$ km s$^{-1}$), both of which are fundamental parameters for simulating the orbital mechanics of Galactic objects, as previously provided in Section \ref{sec4} Moreover, the vertical position of the Sun with respect to the Galactic plane is assumed to be $Z_{\rm 0} = 25 \pm 5$ pc, contributing to a more accurate depiction of the Galactic structure \citep{Juric_2008}. This modelling approach permits the simulation of star cluster orbits within a realistic Galactic potential, thereby enhancing the precision of dynamical studies in both the disc and halo regions.

\begin{table}
\centering
\small
\renewcommand{\arraystretch}{1.1}
\caption{Dynamic orbit parameters of OCSN.}
\begin{tabular}{lcccc}
\hline
\textbf{Parameter} & \textbf{OCSN 203} & \textbf{OCSN 213} & \textbf{OCSN 244} \\
\hline
\hline
$Z_{\rm max}$ (pc) & 85 $\pm$ 1.00  & 152 $\pm$ 14.00 &  150 $\pm$ 8.00 \\
$R_{\rm a}$ (pc) & 9017 $\pm$ 3.00 & 8513 $\pm$ 295   & 8674 $\pm$ 59.00 \\
$R_{\rm p}$ (pc) & 8203 $\pm$ 2.00 & 8316 $\pm$ 257   & 8027 $\pm$ 37.00 \\
$R_{\rm m}$ (pc) & 8610 $\pm$ 3.00 & 8414 $\pm$ 276   & 8350 $\pm$ 48.00 \\
$e$              & 0.05 $\pm$ 0.01 & 0.01 $\pm$ 0.01  & 0.04 $\pm$ 0.01 \\
$T_{\rm p}$ (Myr) & 242 $\pm$ 1.00  & 236 $\pm$ 1.00   &  234 $\pm$ 5.00 \\
\hline
\end{tabular}
\label{tab6}
\end{table}

\begin{figure*}
\centering
\includegraphics[width=0.99\linewidth]{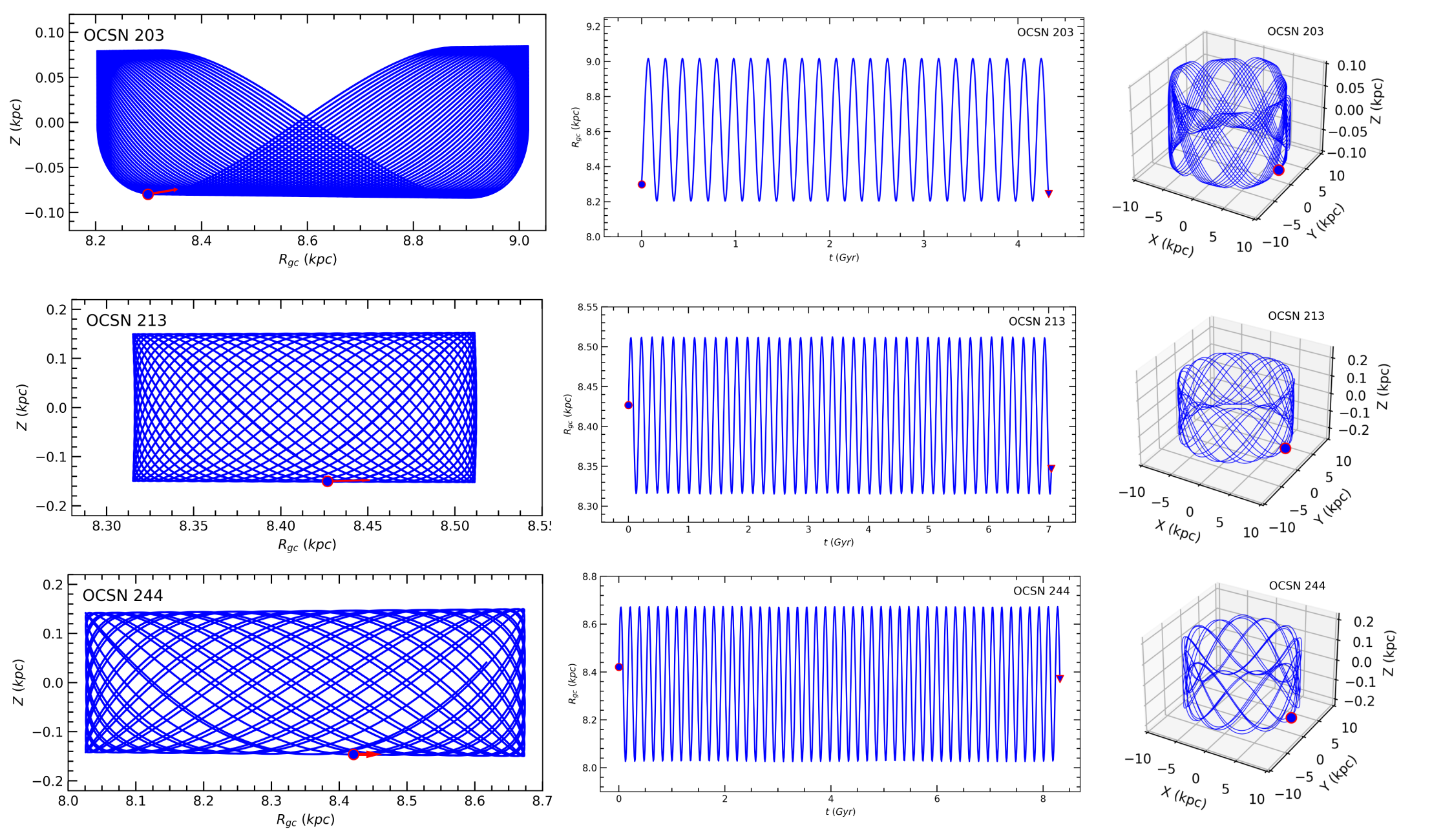}
\caption{The representation of OCSN's Galactic orbits and birth radii is provided on three distinct planes: $Z$ × $R_{\rm gc}$ (left), $R_{\rm gc}$ × $t$  (middle), and $X$ × $Y$ × $Z$ (right). Red circles indicate the current positions of the OCSN, while triangles represent their initial formation locations. Red arrows show the movement vectors of the clusters.}
\label{fig8}
\end{figure*}

To perform a detailed dynamic orbit analysis for the OCSN, several input parameters are required, including quatorial coordinates ($\alpha$, $\delta$), distance ($d$), and mean proper motion components ($\mu_\alpha\cos\delta,~\mu_\delta$). In addition, radial velocity data ($V_{\rm r}$) are required to complete the dynamic orbit analysis. To determine the mean radial velocities of the OCSN clusters, we used the most probable ($P\geq50\%$) cluster members  defined from the $Gaia$ DR3 catalogue and mentioned in $\S$ \ref{sec:astro_parmeters}. The methodology employed is in accordance with the equations summarised by \cite{Soubiran18}, which calculate the mean radial velocity. Numerically the radial velocity $V_{\rm r}$ of these OCs are, 15.97 $ \pm$ 1.69 km s$^{-1}$ (OCSN 203), 26.96 $\pm$ 12.05 km s$^{-1}$ (OCSN 213), and 32.55 $\pm$ 0.43 km s$^{-1}$ (OCSN 244). These findings are consistent with the margins of error of radial velocity estimates reported by \citet{Qin23}. These parameters were utilised to calculate the current positions of the clusters OCSN 203, OCSN 213, and OCSN 244. The dynamic orbits of these clusters were extrapolated forward in time by integrating their dynamic orbits with time steps of 1.00 Myr, corresponding to the age of each OC \citep{Dursun_2024}. The results of the orbital integration are presented in Table \ref{tab6}. As can be seen in Figure \ref{fig8}, OCSN clusters follow nearly circular orbits and considering their maximum height above the Galactic plane, they belong to the young stellar disk component of the MW \citep{Ak_2015}.

\section{Conclusion}
\label{sec7}
In this work, we carried out the first comprehensive study to conduct an exhausting astrometric, photometric, and kinematics of three newly discovered OCs with candidates via a recent and much larger sample of open clusters data adopted today with $Gaia$ DR3 \citep{2021A&A...649A...1G}. Our main findings can be summarized as follows:

\begin{itemize}
\item 	The cluster centers were re-estimated, and the obtained results agree with those obtained by \cite{Qin23} in the Ra direction ($\Delta\alpha\le 2^m~04^{s}.00$) and Dec. ($\Delta\delta\le 10^m~52^{s}.20$).

\item For each cluster, we estimate spatial structure (RDPs) to perform the internal structural analysis, like the core ($1.68\le r_{\rm c} \le 10.31$), limiting ($3.55 \le r_{\rm cl} \le 10.46$), and tidal ($5.06 \le r_{\rm t} \le 14.16$) radii in arcmin units, where ($r_{\rm c} < r_{\rm cl} <  r_{\rm t}$).
	
\item The number $N$ of membership probabilities ($P\ge50\%$) with $Gaia$ DR3 with the aid of UPMASK method is estimated, i.e., ($N_{\rm OCSN~203}=227$), ($N_{\rm OCSN~213}=200$), and ($N_{\rm OCSN~224}=551$). Therefore, CMDs of many isochrones fitting with log age (yr) between ($\log t$; 6.52 – 7.05) and metallicities ($Z$; 0.01308 – 0.01413) are constructed to estimate, reddening $E(G_{\rm BP}-G_{\rm RP})$, and the distance modulus ($m-M$). Our estimated distances are 332 $\pm$ 18, 529 $\pm$ 23, and 506 $\pm$ 23 (pc) for OCSN 203, OCSN 213, and OCSN 244, respectively.
	
\item The estimated LF and with adopted isochrones \citep{Evans18}, make us infer the MLR of individual member stars. Therefore, the collective masses $M_{\rm C}$ of the clusters are given as; 67 $\pm$ 8.19, 91 $\pm$ 9.54, and 353 $\pm$ 18.79 ($M_{\odot}$) OCSN 203, OCSN 213, and OCSN 244, respectively. Both $M_{\rm C}$ and the tidal radius $r_{\rm t}$ reflect us to compute the escape velocity $V_{esc}$ such as 11671 $\pm$ 108, 1999 $\pm$ 45, and 5907 $\pm$ 77 (Km s$^{-1}$) OCSN 203, OCSN 213, and OCSN 244, respectively. On the other hand, the mass function MF slopes are determined (i.e., $\alpha_{\rm OCSN~203}= 2.41 \pm 0.06$, $\alpha_{\rm OCSN~213}= 2.13 \pm 0.07$, and $\alpha_{\rm OCSN~224}= 2.28 \pm 0.07$) which are in good agreement with the value of 2.35 given by \cite{Salpeter55} for masses $\ge 1 M_{\odot}$. 
We also determined the Jacobi radii $r_{\rm J}$ for each cluster, which provide a more reliable estimate than the tidal radii for assessing a cluster's gravitational binding. The computed $r_{\rm J}$ values (5.985 $\pm$ 2.45, 6.658 $\pm$ 2.58, and 10.444 $\pm$ 3.23 pc for OCSN 203, OCSN 213, and OCSN 244, respectively) reflect the mass-loss potential for these clusters throughout their orbits. This highlights that OCSN 213, with its low density contrast parameter ($\delta_{\rm c}$), appears more scattered and less gravitationally bound compared to the others.
	
\item We report that the clusters OCSN 203 and OCSN 244 are dynamically relaxed where the dynamical evolution parameter ($\tau=age/T_{\rm relax}$) is greater than one (i.e., $\tau \gg 1$), while OCSN 213 are non-relaxed cluster.

\item We show their spatial velocity distribution plots along Galactic coordinates and their mean values (i.e., $\overline{U},~\overline{V},~\overline{W}$; km s$^{-1}$). Moreover, their convergent points ($A_{\rm o},~D_{\rm o}$) = (76$^o$.77 $\pm$ 0$^o$.01, -0$^o$.23 $\pm$ 0$^o$.00; OCSN 203), (85$^o$.71 $\pm$ 0$^o$.11, -9$^o$.63 $\pm$ 0$^o$.03; OCSN 213), and (88$^o$.19 $\pm$ 0$^o$.11, -4$^o$.04 $\pm$ 0$^o$.01; OCSN 244) based on AD-diagrams method and the distribution of their space velocity (i.e., $U,~V,~W$; km s$^{-1}$) along the Galactic coordinates.

\item Finally, the maximum heights reached by the clusters OCSN 203, OCSN 213 and OCSN 244 from the Galactic plane are 85 $\pm$ 1.00 pc, 152 $\pm$ 14.00 pc and 150  $\pm$ 8.00 pc respectively, indicating that they belong to the young stellar disc component. The orbits of the clusters are almost circular and they are found to be moving away from the Galactic plane with time.

\end{itemize}

\section*{Acknowledgements}
We sincerely thank the anonymous referee for their valuable suggestions, which greatly enhanced the quality of this paper. This work presents results from the European Space Agency (ESA) space mission Gaia. $Gaia$ data are being processed by the $Gaia$ Data Processing and Analysis Consortium (DPAC). Funding for the DPAC is provided by national institutions, in particular, the institutions participating in the $Gaia$ Multi-Lateral Agreement (MLA). The $Gaia$ mission website is \url{https://www.cosmos.esa.int/gaia}. The $Gaia$ archive website is \url{https://archives.esac.esa.int/gaia}. The authors extend their appreciation to the Deanship of Scientific Research at Northern Border University, Arar, KSA for funding this research work through the project number “NBU-FFR-2024-237-05”.

\bibliography{OCSN}

\end{document}